\documentclass[]{article}

\usepackage{color}
\usepackage[usenames,dvipsnames,svgnames,table]{xcolor}
\usepackage{graphicx}

\small

\def\farcs{.\!\!^{\prime\prime}}
\newcommand{\lesssim}{\raisebox{-0.13cm}{~\shortstack{$<$ \\[-0.07cm]
      $\sim$}}~}

\begin{document}

\twocolumn

\begin{center}
\fboxrule0.02cm
\fboxsep0.4cm
\fcolorbox{Brown}{Ivory}{\rule[-0.9cm]{0.0cm}{1.8cm}{\parbox{7.8cm}
{ \begin{center}
{\Large\em Perspective}

\vspace{0.5cm}

{\Large\bf Disks Around O-type Young Stellar Objects}

\vspace{0.2cm}

{\large\em Maite Beltr\'an}


\vspace{0.5cm}

\centering
\includegraphics[width=0.23\textwidth]{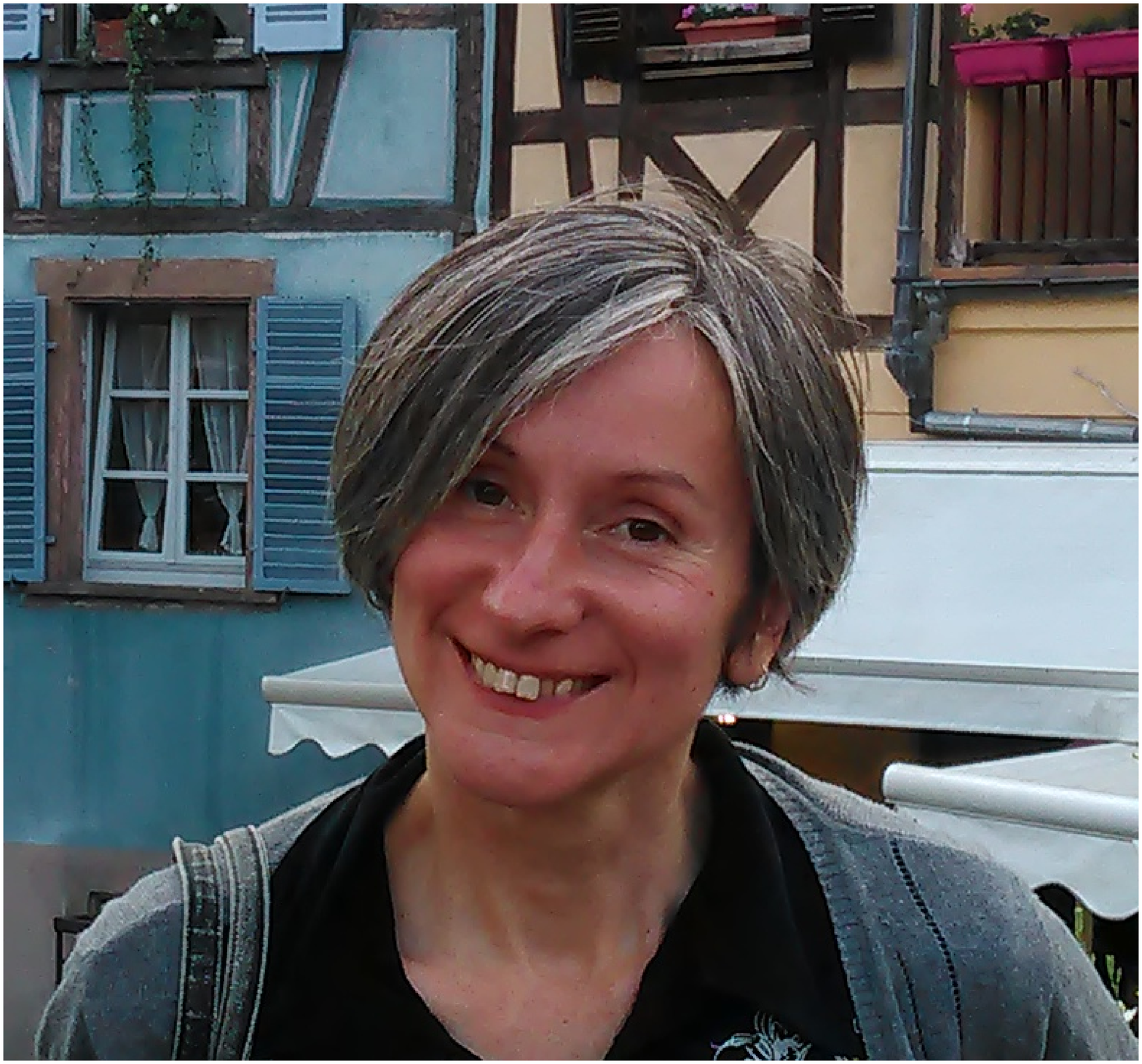}
\end{center}
}}}
\end{center}
 
Accretion disks are one of the key ingredients of the star formation process. They redistribute angular momentum and, in the case of high-mass stars ($M> 8\,M_\odot $), disks would relieve the radiation pressure on the accreting material, in particular in the equatorial direction, by beaming the radiation through the poles of the system and this would allow the accretion to proceed onto the central protostar  (e.g., Tan et al.\ 2014 for a review on massive star formation). In fact, in recent years, all high-mass star-forming theories appear to converge to a disk-mediated accretion scenario (e.g., Krumholz et al.\ 2007; Kuiper et al.\ 2011; Bonnell \& Bate 2006; Keto 2007) but do the observations of high-mass young stellar objects (YSOs) confirm the theory predictions? Or in other words, do true accretion disks around massive stars really exist?

\begin{figure}[ht!]
\centering
\includegraphics[width=0.49\textwidth]{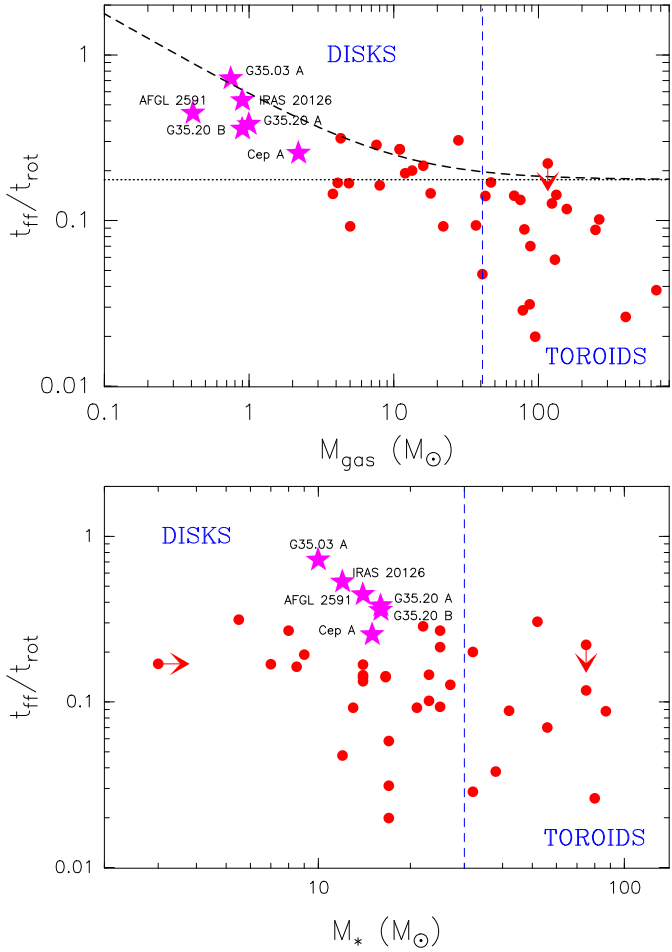}
\caption{({\it Upper panel}) Figure adapted from Beltr\'an \& de Wit (2016). Free-fall timescale to rotational period ratio, $t_{\rm ff}/t_{\rm rot}$, versus $M_{\rm gas}$, of rotating disks or toroids around high-mass (proto)stars. The magenta stars and the labels correspond to the most likely Keplerian disk candidates around B-type (proto)stars. Black dotted and dashed lines correspond to the theoretical values of $t_{\rm ff}/t_{\rm rot}$ for spherical clouds of mass $M_{\rm gas}$ containing a star of mass $M_{\star}$ at the center, in which the gas is rotationally supported against the gravity of both the gas plus the star. The dotted black line corresponds to $M_{\star}=0\,M_\odot$ and the dashed one to $10\,M_\odot$. The blue dashed line indicates a $M_{\rm gas}$ of $40\,M_\odot$. The rotating structures with masses higher than this value are toroids. ({\it Lower panel}) Same as above but for mass of the central star, $M_{\star}$, instead. $M_{\star}$ corresponds to $M_{\star\, \rm cluster}$ of Table 1 of  Beltr\'an \& de Wit (2016), which is the mass of the most massive star in the simulated cluster. The blue dashed line, in this case, indicates a $M_\star$ of $30\,M_\odot$.}
\label{fig:fig1}
\end{figure}

In 2016, Willem-Jan de Wit and myself wrote a review on accretion disks in luminous YSOs to try to answer such questions (Beltr\'an \& de Wit 2016). We concluded that the clear signatures of rotating and/or accretion disks reported in the literature  confirmed the existence of circumstellar disks around stars with masses up to 20--30\,$M_\odot$ or $\sim 10^5\,L_\odot$, which would correspond to early B-type or late O-type stars (see Fig.~\ref{fig:fig1}). The disks of these sources have been spatially resolved using line and continuum tracers from infrared to centimeter wavelengths, and maser emission  (e.g., IRAS 20126+4104: Cesaroni et al., 2005, 2014; Cepheus A HW2: Patel et al., 2005; IRAS 13481$-$6214: Kraus et al., 2010; CRL 2136: de Wit et al., 2011), and their kinematics appears to  be consistent, for most cases, with Keplerian rotation (e.g., Bik \& Thi, 2004; Blum et al.\ 2004: Cesaroni et al.\ 2005, 2014; Wheelwright et al.\ 2010; Wang et al.\ 2012; Ilee et al.\ 2013; S\'anchez-Monge et al.\ 2013; Beltr\'an et al.\ 2014). As a result of our study, we also concluded that the accumulated evidence for disks in young OB-type stars did not extend to stars beyond a mass limit of $\sim$30\,$M_\odot$, that is, to early O-type stars. Instead, what had been observed towards these sources were more massive ($> 100\,M_\odot$) and larger ($> 10^3$\,au) rotating structures called toroids. When we published the study, the ALMA era had just started, and up to then, no high-angular resolution ALMA observations of O-type disk candidates had been yet reported (note that the work by Johnston et al.\ 2015 appeared when our review was already in press).  More than five years later, ALMA has reached an almost complete configuration and long-baseline observations have become available, so, it is time to reformulate our original question and this time confirm or discard the existence of true accretion disks around O-type young stars,, in particular around early O-type stars. If confirmed, this would mean that accretion disks are essential for the formation of stars of all luminosities. 
 
{\bf ALMA observations of O-type YSOs}

-- \underline{O-type YSOs with $L_{\rm bol}>10^5\,L_\odot$}

Johnston et al.\ (2015) reported the first candidate disk around an O-type (proto)star of 
$\sim2\times10^5\,L_\odot$, AFGL\,4176, observed with ALMA.  For this source, the emission of CH$_3$CN, a typical high-density tracer, shows a velocity gradient along the major axis of the source that was consistent with  Keplerian-like rotation. The position-velocity (PV) plots show a clear ``butterfly" pattern with low-velocity  ``spurs'' and high-velocity spikes close to the position of the embedded protostar, that is consistent with the pattern expected for an edge-on Keplerian disk rotating about a 25\,$M_\odot$  O7 star. 

Cesaroni et al.\ (2017) carried out ALMA observations with an angular resolution of $\sim0\farcs2$ of a small sample of six star-forming regions containing the most luminous ($L_{\rm bol}>10^5\,L_\odot$) O-type sources. The observations spatially resolved the CH$_3$CN emission in seven cores and found evidence of rotation for six of them (G24.78$+$0.08 A1, G29.96$-$0.02 HMC, G345.49+1.47 M, G345.50+0.35 M and G345.50+0.35 S). However, only three of them (G29.96$-$0.02 HMC, G345.50+0.35 M and G345.50+0.35 S) show evidence of Keplerian-like rotation. For one source, G17.64+0.16, these observations did not find any evidence of rotation.  Cesaroni et al.\ (2017) plotted the luminosity-to-mass ratio ($L_{\rm bol}/M_{\rm gas}$), which is an evolutionary  stage indicator, versus the distance to the source, including the O-type star AFGL4176, and concluded that the true accretion disk detection rate could be sensitive to the evolutionary stage of the young stellar object. For the youngest sources, the non-detection of Keplerian disks could be due to the fact that the emission of disks is difficult to disentangle from that of the envelopes. Alternatively, disks might start small and grow up with time. On the other hand, for the most evolved source in the sample (G17.64+0.16), the molecular gas might have already been dispersed and therefore, no disk is found. This study concluded that only sources with an intermediate evolutionary stage appeared to have true accretion disks clearly detectable in typical hot core tracers. 

In the meanwhile, since the publication of the Cesaroni et al.\ study, things have slightly changed, in particular for G17.64+0.16, the source for which no disk evidence was found. Maud et al.\ (2018) analyzed and modeled the ALMA SiO observations at $0\farcs2$, also covered by the frequency setup of the observations of Cesaroni et al.\ (2017), and found evidence for a disk in Keplerian rotation around a 10--$30\,M_\odot$ YSO. New ALMA observations of H$_2$O at an even higher angular resolution ($\sim$17\,mas, which corresponds to a spatial resolution of $\sim$38\,au), have recently confirmed the existence of a Keplerian disk around a $45\pm10\,M_\odot$ O-type (proto)star (Maud et al. 2019). Another source for which things have changed is G31.41+0.31. In this case, new parallax observations (Reid et al.\ 2019) have located this high-mass star-forming region much closer, at 3.7 kpc. With this new distance estimate, the luminosity of the source would be of $5.7\times10^4\,L_\odot$,  and therefore, it would not fulfill one of the criterion selection of the Cesaroni et al.\ (2017) sample, that is, to have luminosity $>10^5 L_\odot$.  
-- \underline{O-type YSOs with $L_{\rm bol}<10^5\,L_\odot$}

Moving to lower luminosity ($L_{\rm bol}<10^5\,L_\odot$) O-type YSOs, ALMA observations have revealed new disk candidates. Ilee et al.\ (2016) carried out  millimeter sub-arcsecond SMA observations of the young proto-O star G11.92$-$0.61 MM1, which is embedded in an infrared dark cloud, and discovered a velocity gradient consistent with Keplerian rotation about a 30--$60\,M_\odot$ YSO. The central mass modeled by Ilee et al. (2016) is much higher than the one that would correspond to a luminosity of $10^4\,L_\odot$. Following Beltr\'an \& de Wit (2016) and assuming that $L_{\rm bol}$ is consistent with that of a stellar cluster populated  assuming a randomly sampled Chabrier (2005) initial mass function, we estimated that the mass of the most massive star in the simulated cluster would be of 12$\,M_\odot$ for such a luminosity. This low luminosity could be due to a high accretion rate in this very young high-mass protostar that would increase the radius of the source and decrease its effective temperature (e.g., Hosokawa \& Omukai 2009). Another explanation could be that the SMA observations do not provide high enough angular resolution to properly resolve and model the velocity field around G11.92$-$0.61 MM1. Ilee et al.\ (2018) observed the source with ALMA at about five times higher angular resolution and confirm the presence of a disk in Keplerian rotation. The velocity field was fitted with a model of a thin Keplerian disk rotating about a 34--38$\,M_\odot$ central YSO, confirming that  G11.92$-$0.61 MM1 is one of the most massive O-type Keplerian disk candidates to date.  

Other disk candidates recently discovered with ALMA are G23.01$-$0.41 (Sanna et al.\ 2019) and IRAS\,16547$-$4247 (Zapata et al.\ 2019), both surrounding a central star of $\sim$20\,$M_\odot$.  In the first case, Sanna et al.\ (2019) have reported sub-Keplerian velocities, while in the latter case the spatially resolved asymmetric disk observed by Zapata et al.\ (2019) would be Keplerian. 

On the other hand, several authors have also reported the existence of rotating disk-like structures with velocity gradients that are not (fully) consistent with Keplerian rotation.  In particular, Beuther et al.\ (2017a) observed in CH$_3$CN the $1.7\times10^4\,L_\odot$ O-type YSO G351.77$-$0.54  with ALMA at an angular resolution of only $60\,$mas ($\sim$130\,au) and concluded that although the line emission shows hints of Keplerian rotation, to properly model the velocity field in this region one should take into account also the contribution of infalling and outflowing material.  In a similar way, Maud et al.\ (2017) observed, also in CH$_3$CN, the W33A MM1-Main O-type YSO with a spatial resolution of $\sim$500\,au, and found that the emission showed a very complex morphology with a filamentary and spiralling structure that could feed an embedded and not resolved  circumstellar disk. After analyzing the velocity field, these authors concluded that Keplerian rotation alone cannot satisfactorily reproduce the kinematic signatures, although the possibility of the existence of a small ($<$500\,au) Keplerian disk embedded in the core cannot be discarded.  Finally, although the observations were not carried out with ALMA but with the VLA, it is worth to mention the case of the $8\times10^4\,L_\odot$ O-type YSO NGC\,7538IRS1 (Beuther et al.\ 2017b). VLA CH$_3$OH observations at $\sim$150\,au spatial resolution revealed two disk-like structures with velocity gradients consistent with rotation but with no Keplerian signatures. Beuther et al.\ (2017b)  concluded that taking into account the early evolutionary stage of NGC\,7538IRS1, which is still undergoing a major accretion phase as indicated by redshifted absorption, the Keplerian disks could still be too small ($<190$\,au) to be detected and this would be in line with the simulations of Kuiper et al.\ (2011). In addition, this material should be ionized, as indicated by the presence of hypercompact HII regions and therefore, the thermal emission of CH$_3$OH could not be the best tracer of the accreting and rotating putative Keplerian disk. 

It is worth to mention here that, as recently demonstrated by Ahmadi et al.\ (2019), one has to be cautious when ruling out the existence of true accretion disks based on the appearance of the PV plots because Keplerian disks can mimic solid-body rotation when observed with poor angular resolution. 

In summary, in recent years very high-angular resolution ALMA (and VLA) observations have revealed velocity gradients consistent with rotation at disk scales in thirteen star-forming regions associated with O-type YSOs, with six of them with luminosity $>10^5\,L_\odot$, and seven with $<10^5\,L_\odot$. In a couple of regions, G345.50+0.35 and NGC\,7538IRS1, the observations have resolved the line emission in more than one rotating structure. Therefore, the number of rotating disk-like structures is seven for $L_{\rm bol}>10^5\,L_\odot$ and eight for $L_{\rm bol}<10^5\,L_\odot$. Finally, five out of seven structures show Keplerian rotation signatures for sources with $L_{\rm bol}>10^5\,L_\odot$, and three out of eight for sources with $L_{\rm bol}<10^5\,L_\odot$.

{\bf Disk kinematics}

\begin{figure}[ht!]
\centering
\includegraphics[width=0.49\textwidth]{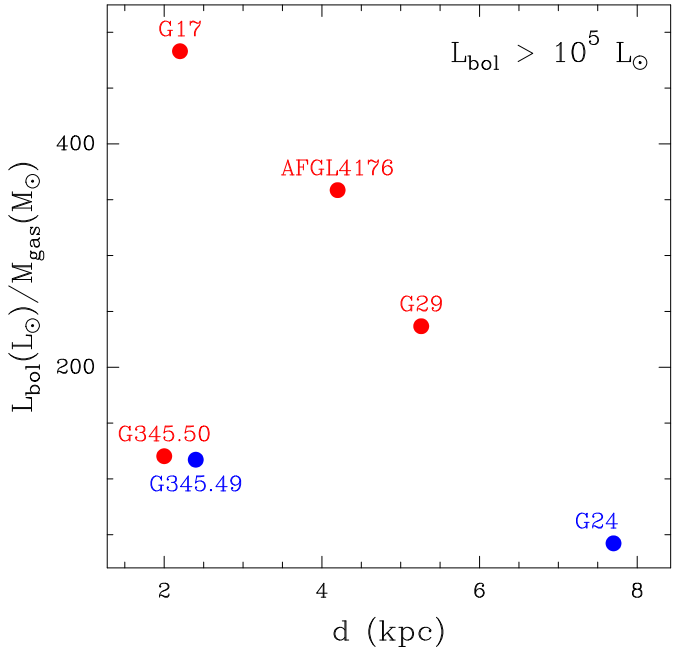}
\caption{Luminosity-to-mass ratio versus source distance for high-mass star-forming regions with $L_{\rm bol}>10^5\,L_\odot$. Red and blue points indicate sources with disk candidates (Keplerian rotation) and with questionable disk candidates (no Keplerian rotation evidence), respectively.}
\label{fig:fig2}
\end{figure}

Following Cesaroni et al.\ (2017), with this new statistics on disk-like structures around O-type YSOs, it is possible to investigate whether the presence of true accretion disks around these massive stars is associated with their evolutionary stage. To do this, we estimated the $L_{\rm bol}/M_{\rm gas}$ ratio for each of the thirteen, where the mass  of the molecular clump, $M_{\rm gas}$ was computed from the ATLASGAL 870$\,\mu$m flux density (Schuller et al.\ 2009) catalogued by Contreras et al.\ (2013), assuming, like Cesaroni et al.\ (2017), a dust temperature $T_{\rm dust}$  of 50\,K, a dust absorption coefficient $\kappa_{\rm dust}$  of 2\,cm$^2$\,g$^{-1}$ at 870\,$\mu$m (Ossenkopf \& Henning 1994), and a gas-to-dust mass ratio of 100.  Note that in our case, the masses have been estimated from the flux density obtained from the ATLASGAL catalogue (Contreras et al.\ 2013), while Cesaroni et al.\ (2017) estimated the flux density of the sources from the ATLASGAL maps. Therefore, the  $L_{\rm bol}/M_{\rm gas}$ values in both cases are slightly different. For NGC\,7538IRS1 for which no ATLASGAL data are available, we have used the fluxes of BOLOCAM at 1.1\,mm (Ginsburg et al.\ 2013), using the same  $T_{\rm dust}$ of 50\,K  and a dust absorption coefficient $\kappa_{\rm dust}$  of 1.24\,cm$^2$\,g$^{-1}$.   
Figure~\ref{fig:fig2} is an update of Fig.~29 of Cesaroni et al.\ (2017), where we plot the $L_{\rm bol}/M_{\rm gas}$ ratio as a function of distance for sources with $L_{\rm bol}>10^5\,L_\odot$. We have removed G31.41+0.31 from the plot because, with the new distance estimate, its luminosity is of 5.7$\times10^4\,L_\odot$ and have changed the "status" to G17.64+0.16, after the discovery of a disk with Keplerian rotation signatures by Maud et al.\ (2018). Unfortunately, no new disk candidates with luminosities $>10^5\,L_\odot$ have been discovered since the study of Cesaroni et al.\ (2017). As seen in Fig.~\ref{fig:fig2}, the evolutionary trend observed by Cesaroni et al.\ still holds; that is, O-type can be surrounded by true accretion disks, and the fact that such disks are not detected for all of them is because the youngest sources are too deeply embedded to properly disentangle the velocity field of the disk from that of the surrounding envelope or toroid. 

\begin{figure}[ht!]
\centering
\includegraphics[width=0.49\textwidth]{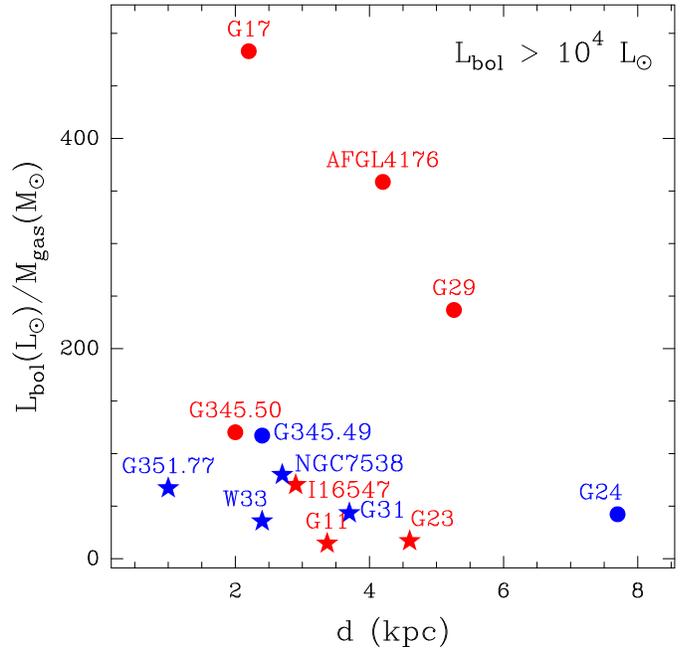}
\caption{Same as Fig.~\ref{fig:fig2} for high-mass star-forming regions with $L_{\rm bol}>10^5\,L_\odot$ ({\it dots}) and $L_{\rm bol}<10^5\,L_\odot$ ({\it stars}). Red and blue symbols indicate sources with disk candidates (Keplerian rotation) and with questionable disk candidates (no Keplerian rotation evidence), respectively.}
\label{fig:fig3}
\end{figure}

To check whether this trend is still valid if one takes into account all O-type sources for which velocity gradients suggestive of rotation have been observed, in Fig.~\ref{fig:fig3} we have plotted the $L_{\rm bol}/M_{\rm gas}$ ratio as a function of distance but this time for sources with $L_{\rm bol}>10^4\,L_\odot$. As seen in Fig.~\ref{fig:fig3}, when taking into account all the O-type sources (early and late) the evolutionary trend disappears and there is no more correlation between the youth of the YSOs and their lack of accretion disks. In fact, the youngest YSOs, with $L_{\rm bol}/M_{\rm gas}\sim 15$, have Keplerian disks around them. However, while for early O-type YSOs, the $L_{\rm bol}/M_{\rm gas}$ ratio spans more than one order of magnitude, for late O-type sources the range of this ratio is very limited and only ranges from $\sim$15 to 80. Therefore, it is more difficult to clearly establish a difference in the evolutionary stage of the sources.  In any case, a possible explanation of the fact that Keplerian disks were detected around the younger sources for late O-type YSOs but were not detected for early O-type ones could be due to the fact that the former objects have less material around them and therefore, it is possible to disentangle the surrounding envelope from the disk.

That no Keplerian disks have been found for some late O-type YSOs does not mean that true accretion disks do not exist in these sources. In fact, recent observations suggest that Keplerian accretion disks maybe small and hidden on scales $< 500$\,au (e.g., W33A: Maud et al.\ 2017;  NGC 7538IRS1: Beuther et al.\ 2017b). This would be supported by models and simulations that predict that Keplerian disks, constantly replenished by infalling and rotating material from large-scale non-Keplerian structures, would grow with time, from the inside outward, from $\sim$10\,au up to $>$1000\,au (Kuiper et al.\ 2011).   

{\bf Disk substructure and stability}

Interestingly,  very high angular resolution observations of some Keplerian disk candidates have revealed the existence of substructure in the disk, in the form of spirals (G23.01$-$0.41: Sanna et al.\ 2019;  AFGL\,4176: Johnston et al.\ 2020) and rings (G17.64+0.16: Maud et al.\ 2019). ALMA observations at a spatial resolution of $\sim$500\,au have also revealed a spiral arm in W33A, for which no Keplerian rotation signature has been found (Maud et al.\ 2017). Spiral arms feeding the central star could develop by the growth of eccentric gravitational instabilities in the star/disk system (the SLING mechanism of Adams et al.\ 1989) and have been predicted by numerical simulations of disk formation around high-mass YSOs (e.g., Krumholz et al.\ 2007; Kuiper et al.\ 2011; Ahmadi et al.\ 2019). The same simulations show that, at larger scales, spiral structures channel material from the surrounding large-scale infalling envelope onto the embedded circumstellar disk. The latter would be the case of W33A, where the spiral arm observed has been interpreted as an accreting filament that would feed a likely embedded disk with material from the large-scale infalling envelope (Maud et al.\ 2017).

The stability of the disks against gravitational instabilities can be estimated by means of the Toomre $Q$ parameter (Toomre 1964). According to this stability criterion, a thin disk becomes unstable against axisymmetric gravitational instabilities and prone to fragmentation if $Q < 1$, while non-axisymmetric instabilities, which grow as multi-armed spirals, become unstable for values of $Q$ slightly higher, between 1 and 2 (e.g., Papaloizou \& Savonije 1991; Durisen et al.\ 2007).  For AFGL\,4176, Johnston et al.\ (2020) have estimated $Q$ across the disk and have found that the spiral arms are Toomre unstable, with $Q<2$, and therefore, they could undergo fragmentation. For the other exhibiting source spiral arms, G23.01$-$0.41, no study of the stability of the disk has been carried out. For the other Keplerian disk candidate showing structure, G17.64+0.16, Maud et al.\ (2019) conducted a Toomre $Q$ stability analysis, and found the disk, and in particular the rings, to be stable against fragmentation ($Q>2$). These authors propose that the substructures observed in the disk (the rings) might have formed from fragmentation in earlier unstable phase. Regarding the other O-type YSOs with rotating structures, a Toomre $Q$ parameter stability study has only been performed for G351.77$-$0.54 by Beuther et al.\ (2017a), who, having found $Q>1.5$, have concluded that the rotating structure should be stable against axisymmetric gravitational fragmentation. 

The stability of the rotating structures can also be investigated via the ratio of the mass of the rotating structure and the mass of the central star, $M_{\rm rot}/M_\star$. According to theory, for $M_{\rm rot}/M_\star>3$, gravitational instabilities that induce spiral density waves appear, leading to a rapid fragmentation of the disk (Laughlin \& Bodenheimer 1994; Yorke 1995). We have compiled the $M_{\rm rot}/M_\star$ for those objects for which it has been possible to find an estimate of the mass of the rotating structure, nine out of fifteen. For all but one case (G31.41+0.31), we have found that $M_{\rm rot}/M_\star\lesssim3$, which would suggest that almost all the rotating structures should be stable. However, because all the observations have been carried out with interferometers at very high-angular resolution, it cannot be discarded the possibility that some emission of the rotating structure has been filtered out, and therefore,  $M_{\rm rot}$ should be taken as a lower limit.

{\bf Conclusions}

Accretion disks around high-mass YSOs are fundamental to understand how such stars form. The existence of true accretion disks has been been well established for stars with masses up to $\sim$20$\,M_\odot$, which correspond to early B-type or late O-type stars, thanks to the unprecedented angular resolution and sensitivity provided by ALMA in recent years. However, despite the progress made in this field, the number of early O-type (proto)stars ($L>10^5\,L_\odot$) with Keplerian rotation signatures is still very limited. Up to now, the best true disk candidates are AFLG\,4176 and G17.64+0.16 that host a Keplerian disk rotating about a 25\,$M_\odot$ and a 45\,$M_\odot$ star, respectively (Johnston et al.\ 2015, 2020; Maud et al.\ 2018, 2019). G11.92$-$0.61\,MM1 also deserves special mention because despite having a luminosity of only $10^4\,L_\odot$, this source could be hosting a Keplerian disk rotating about a 34--38 $M_\odot$  YSO. The fact that the luminosity of this object is so low is because G11.92$-$0.61\,MM1 is a proto-O star and is still deeply embedded in the parental cloud. These cases are encouraging, but the numbers are still too small to establish, on statistical grounds, that stars of all luminosities are formed via disk-mediated accretion. However, since the Keplerian disk detections around early O-type stars are very recent, the situation is bound to change, and we expect more confirmed cases in the coming years. In particular,  when the future baseline expansion of ALMA (aiming at an angular resolution of 0$\farcs$001--0$\farcs$003) becomes a reality.

\normalsize

\footnotesize

{\bf References:}\\

Adams, F.\ C., Ruden, S.\ P., \& Shu, F.\ H.\ 1989,  ApJ, 347, 959\\
Ahmadi, A., Kuiper, R., Beuther, H.\  2019, A\&A, 632, A50 \\
Beltr\'an, M.\ T., \& de Wit 2016, A\&ARv, 24, 6 \\
Beltr\'an, M.\ T., et al.\ 2014, A\&A, 571, A52 \\
Beuther, H., et al.\ 2017a, A\&A, 603, A10 \\
Beuther, H., et al.\ 2017b, A\&A, 605, A61 \\
Bik, A., \& Thi, W.\ 2004, A\&A, 427, L13 \\
Blum, R., et al.\ 2004, ApJ, 617, 1167 \\ 
Bonnell, I., \& Bate, M.\ 2006, MNRAS, 370, 488 \\
Cesaroni, R., et al.\ 2005, A\&A, 434, 1039 \\
Cesaroni, R., et al.\ 2014, A\&A, 566, A73 \\
Cesaroni, R., et al.\ 2017, A\&A, 609, A59 \\
Chabrier, G.\ 2005, in Initial Mass Function 50 Years Later, vol.\ 327 of Astrophysics and Space Science Library,  41 \\
Contreras, Y., et al.\ 2013, A\&A 549, A45 \\ 
Durisen, R.\ H.,  et al.\ 2007, in Protostars \& Planets V, 607  \\
Hosokawa, T., \& Omukai, K.\ 2009, ApJ, 691, 823 \\
Ilee, J.\ D., et al.\ 2013, MNRAS, 429, 2960 \\
Ilee, J.\ D., et al.\  2016, MNRAS, 462, 4386 \\ 
Ilee, J.\ D., et al.\ 2018, ApJ, 869, L24 \\ 
Johnston, K.\ G., et al.\ 2015, ApJL, 813, L19 \\
Johnston, K.\ G., et al.\ 2020, A\&A, 634, L11  \\
Keto, E.\ 2007, ApJ, 666, 976 \\
Krumholz, M.\ R., Klein, R.\ I., \& McKee, C.\ F.\ 2007, ApJ, 665, 478 \\
Kuiper, R., et al.\  2011, ApJ, 732, 20 \\
Laughlin, G., \& Bodenheimer, P.\ 1994, ApJ, 436, 335 \\
Maud, L.\ T., et al.\ 2017, MNRAS, 467, L120 \\
Maud, L.\ T., et al.\ 2018, A\&A, 620, A31 \\
Maud, L.\ T., et al.\ 2019, A\&A, 627, L6 \\
Papaloizou, J.\ C., \& Savonije, G.\ J.\ 1991, MNRAS, 248, 353 \\ 
Patel, N., et al.\ 2005, Nature, 437, 109 \\ 
Reid, M., et al.\ 2019, ApJ, 885, 131 \\ 
S\'anchez-Monge, \'A., et al.\ 2013, A\&A, 552, L10 \\ 
Sanna, A., et al.\ 2019, A\&A, 623, A77 \\
Schuller, F., et al.\ 2009, A\&A, 504, 415 \\
Tan, J.\ C., et al.\ 2014, in Protostars and Planets, VI, 149 \\
Toomre, A.\ 1964, ApJ, 139, 1217 \\
Wang, K.-S., van der Tak, F.\ F.\ S., \& Hogerheijde, M.\ R.\ 2012, A\&A, 543, A22 \\
Wheelwright, H., et al.\ 2010, MNRAS, 408, 1840 \\
Yorke, H.\ 1995, RevMexAA, Conference Series 1, 35 \\
Zapata, L.\ A., et al.\ 2019, ApJ, 872, 176 \\

\normalsize

\end{document}